\documentstyle[twoside]{article}

\catcode`\@=11
\long\def\@makefntext#1{
\protect\noindent \hbox to 3.2pt {\hskip-.9pt
$^{{\eightrm\@thefnmark}}$\hfil}#1\hfill}		

\def\thefootnote{\fnsymbol{footnote}}
\def\@makefnmark{\hbox to 0pt{$^{\@thefnmark}$\hss}}	

\def\ps@myheadings{\let\@mkboth\@gobbletwo
\def\@oddhead{\hbox{}
\rightmark\hfil\eightrm\thepage}
\def\@oddfoot{}\def\@evenhead{\eightrm\thepage\hfil
\leftmark\hbox{}}\def\@evenfoot{}
\def\sectionmark##1{}\def\subsectionmark##1{}}

\oddsidemargin=\evensidemargin
\renewcommand{\thefootnote}{\fnsymbol{footnote}}

\newcounter{sectionc}\newcounter{subsectionc}\newcounter{subsubsectionc}
\renewcommand{\section}[1] {\vspace{12pt}\addtocounter{sectionc}{1}
\setcounter{subsectionc}{0}\setcounter{subsubsectionc}{0}\noindent
	{\tenbf\thesectionc. #1}\par\vspace{5pt}}
\renewcommand{\subsection}[1] {\vspace{12pt}\addtocounter{subsectionc}{1}
	\setcounter{subsubsectionc}{0}\noindent
	{\bf\thesectionc.\thesubsectionc. {\kern1pt \bfit #1}}\par\vspace{5pt}}
\renewcommand{\subsubsection}[1] {\vspace{12pt}\addtocounter{subsubsectionc}{1}
	\noindent{\tenrm\thesectionc.\thesubsectionc.\thesubsubsectionc.
	{\kern1pt \tenit #1}}\par\vspace{5pt}}
\newcommand{\nonumsection}[1] {\vspace{12pt}\noindent{\tenbf #1}
	\par\vspace{5pt}}

\newcounter{appendixc}
\newcounter{subappendixc}[appendixc]
\newcounter{subsubappendixc}[subappendixc]
\renewcommand{\thesubappendixc}{\Alph{appendixc}.\arabic{subappendixc}}
\renewcommand{\thesubsubappendixc}
	{\Alph{appendixc}.\arabic{subappendixc}.\arabic{subsubappendixc}}

\renewcommand{\appendix}[1] {\vspace{12pt}
        \refstepcounter{appendixc}
        \setcounter{figure}{0}
        \setcounter{table}{0}
        \setcounter{lemma}{0}
        \setcounter{theorem}{0}
        \setcounter{corollary}{0}
        \setcounter{definition}{0}
        \setcounter{equation}{0}
        \renewcommand{\thefigure}{\Alph{appendixc}.\arabic{figure}}
        \renewcommand{\thetable}{\Alph{appendixc}.\arabic{table}}
        \renewcommand{\theappendixc}{\Alph{appendixc}}
        \renewcommand{\thelemma}{\Alph{appendixc}.\arabic{lemma}}
        \renewcommand{\thetheorem}{\Alph{appendixc}.\arabic{theorem}}
        \renewcommand{\thedefinition}{\Alph{appendixc}.\arabic{definition}}
        \renewcommand{\thecorollary}{\Alph{appendixc}.\arabic{corollary}}
        \renewcommand{\theequation}{\Alph{appendixc}.\arabic{equation}}
        \noindent{\tenbf Appendix \theappendixc #1}\par\vspace{5pt}}
\newcommand{\subappendix}[1] {\vspace{12pt}
        \refstepcounter{subappendixc}
        \noindent{\bf Appendix \thesubappendixc. {\kern1pt \bfit #1}}
	\par\vspace{5pt}}
\newcommand{\subsubappendix}[1] {\vspace{12pt}
        \refstepcounter{subsubappendixc}
        \noindent{\rm Appendix \thesubsubappendixc. {\kern1pt \tenit #1}}
	\par\vspace{5pt}}

\newcommand{\textlineskip}{\baselineskip=13pt}
\newcommand{\smalllineskip}{\baselineskip=10pt}

\def\eightcirc{
\begin{picture}(0,0)
\put(4.4,1.8){\circle{6.5}}
\end{picture}}
\def\eightcopyright{\eightcirc\kern2.7pt\hbox{\eightrm c}}

\newcommand{\copyrightheading}[1]
	{\vspace*{-2.5cm}\smalllineskip{\flushleft
	{\footnotesize International Journal of Modern Physics A, #1}\\
	{\footnotesize $\eightcopyright$\, World Scientific Publishing
	 Company}\\
	 }}

\newcommand{\publisher}[2]{{\begin{center}\footnotesize\smalllineskip
	Received #1\\
	Revised #2
	\end{center}
	}}

\def\abstracts#1#2#3{{
	\centering{\begin{minipage}{4.5in}\baselineskip=10pt\footnotesize
	\parindent=0pt #1\par
	\parindent=15pt #2\par
	\parindent=15pt #3
	\end{minipage}}\par}}

\renewenvironment{thebibliography}[1]
	{\frenchspacing
	 \ninerm\baselineskip=11pt
	 \begin{list}{\arabic{enumi}.}
	{\usecounter{enumi}\setlength{\parsep}{0pt}
	 \setlength{\leftmargin 12.7pt}{\rightmargin 0pt} 
	 \setlength{\itemsep}{0pt} \settowidth
	{\labelwidth}{#1.}\sloppy}}{\end{list}}

\newcounter{itemlistc}
\newcounter{romanlistc}
\newcounter{alphlistc}
\newcounter{arabiclistc}

\newcommand{\fcaption}[1]{
        \refstepcounter{figure}
        \setbox\@tempboxa = \hbox{\footnotesize Fig.~\thefigure. #1}
        \ifdim \wd\@tempboxa > 5in
           {\begin{center}
        \parbox{5in}{\footnotesize\smalllineskip Fig.~\thefigure. #1}
            \end{center}}
        \else
             {\begin{center}
             {\footnotesize Fig.~\thefigure. #1}
              \end{center}}
        \fi}

\newcommand{\tcaption}[1]{
        \refstepcounter{table}
        \setbox\@tempboxa = \hbox{\footnotesize Table~\thetable. #1}
        \ifdim \wd\@tempboxa > 5in
           {\begin{center}
        \parbox{5in}{\footnotesize\smalllineskip Table~\thetable. #1}
            \end{center}}
        \else
             {\begin{center}
             {\footnotesize Table~\thetable. #1}
              \end{center}}
        \fi}

\def\@citex[#1]#2{\if@filesw\immediate\write\@auxout
	{\string\citation{#2}}\fi
\def\@citea{}\@cite{\@for\@citeb:=#2\do
	{\@citea\def\@citea{,}\@ifundefined
	{b@\@citeb}{{\bf ?}\@warning
	{Citation `\@citeb' on page \thepage \space undefined}}
	{\csname b@\@citeb\endcsname}}}{#1}}

\newif\if@cghi
\def\cite{\@cghitrue\@ifnextchar [{\@tempswatrue
	\@citex}{\@tempswafalse\@citex[]}}
\def\citelow{\@cghifalse\@ifnextchar [{\@tempswatrue
	\@citex}{\@tempswafalse\@citex[]}}
\def\@cite#1#2{{$\null^{#1}$\if@tempswa\typeout
	{IJCGA warning: optional citation argument
	ignored: `#2'} \fi}}

\def\pmb#1{\setbox0=\hbox{#1}
	\kern-.025em\copy0\kern-\wd0
	\kern.05em\copy0\kern-\wd0
	\kern-.025em\raise.0433em\box0}

\def\fnt#1#2{\footnotetext{\kern-.3em
	{$^{\mbox{\scriptsize #1}}$}{#2}}}

\def\fpage#1{\begingroup
\voffset=.3in
\thispagestyle{empty}\begin{table}[b]\centerline{\footnotesize #1}
	\end{table}\endgroup}

\def\runninghead#1#2{\pagestyle{myheadings}
\markboth{{\protect\footnotesize\it{\quad #1}}\hfill}
{\hfill{\protect\footnotesize\it{#2\quad}}}}
\font\tenrm=cmr10
\font\tenit=cmti10
\font\tenbf=cmbx10
\font\bfit=cmbxti10 at 10pt
\font\ninerm=cmr9

\font\eightrm=cmr8

\def\qed{\hbox{${\vcenter{\vbox{			
   \hrule height 0.4pt\hbox{\vrule width 0.4pt height 6pt
   \kern5pt\vrule width 0.4pt}\hrule height 0.4pt}}}$}}

\renewcommand{\thefootnote}{\fnsymbol{footnote}}	

\def\D{{\rm D}}
\def\be{\begin{equation}}
\def\ee{\end{equation}}
\def\bea{\begin{eqnarray}}
\def\eea{\end{eqnarray}}
\def\li#1{\hbox{${}^{#1}$Li}}
\def\he#1{\hbox{$^{#1}{\rm He}$}}
\def\la{~\mbox{\raisebox{-.6ex}{$\stackrel{<}{\sim}$}}~}
\def\ga{~\mbox{\raisebox{-.6ex}{$\stackrel{>}{\sim}$}}~}

\begin{document}

\runninghead{Big Bang Nucleosynthesis
$\ldots$} {Big Bang Nucleosynthesis
$\ldots$}

\normalsize\textlineskip
\thispagestyle{empty}
\setcounter{page}{1}

\copyrightheading{}			
\rightline{UMN-TH-1341/95}
\rightline{June 1995}
\vspace*{0.88truein}

\fpage{1}
\centerline{\bf BIG BANG NUCLEOSYNTHESIS: AN UPDATE}
\vspace*{0.37truein}
\centerline{\footnotesize KEITH A. OLIVE}
\vspace*{10pt}
\centerline{\footnotesize AND}
\vspace*{10pt}
\centerline{\footnotesize SEAN T. SCULLY}
\vspace*{0.015truein}
\centerline{\footnotesize\it School of Physics and Astronomy, University
of Minnesota, 116 Church St. SE}
\baselineskip=10pt
\centerline{\footnotesize\it Minneapolis, MN 55455, USA}
\vspace*{0.225truein}
\publisher{(received date)}{(revised date)}

\vspace*{0.21truein}
\abstracts{The current status of big bang nucleosynthesis is reviewed
with an emphasis on the comparison between the observational
determination of the light element abundances of \D, \he3,
\he4 and \li7 and the predictions from theory. In particular,
we present new analyses for \he4 and \li7.
Implications
for physics beyond the standard model are also discussed. Limits
on the effective number of neutrino flavors
are also updated.}{}{}


\vspace*{1pt}\textlineskip	
\section{Introduction}	        
\vspace*{-0.5pt}
\noindent
The overall status of big bang nucleosynthesis is determined by the comparison
of the rather slowly changing theoretical predictions of the light element
abundances and the sometimes quickly changing observationally determined
abundances. The observed elements, \D, \he3, \he4, \li7, have abundances
relative  to hydrogen which span nearly nine orders of magnitude. By and large,
these observations are consistent with the theoretical predictions and
play a key role in determining the consistency of what we refer to as
the standard big bang model and its extrapolation to time scales on the
order of one second.
Here, we will review the status of this consistency.  We begin
by briefly outlining the key sequence of events in the early Universe
which leads to the formation of the light elements and then discuss
the current status of the observations in relation to theory of each of the
light elements.  Finally, we will  discuss the current limits on physics beyond
the standard model.
\pagebreak

\textheight=7.8truein
\setcounter{footnote}{0}
\renewcommand{\thefootnote}{\alph{footnote}}

\section{A Brief Primer on the Theoretical Predictions\cite{rev}}
\noindent
Conditions for the synthesis of the light elements were attained in the
early Universe at temperatures  $T \la $ 1 MeV, corresponding to an age of
about 1 second.  At somewhat higher temperatures, weak interaction rates were
in equilibrium, thus fixing the ratio of number densities of neutrons to
protons. At $T \gg 1$ MeV, $(n/p) \simeq 1$.  As the temperature fell and
approached the point where the weak interaction rates were no longer fast
enough
to maintain equilibrium, the neutron to proton ratio was given approximately by
the Boltzmann factor, $(n/p) \simeq e^{-\Delta m/T}$, where $\Delta m$
is the neutron-proton mass difference. The final abundance of \he4 is very
sensitive to the $(n/p)$ ratio.

The nucleosynthesis chain begins with the formation of deuterium
through the process, ${p+n \rightarrow \D} + \gamma$.
However, because the large number of photons relative to nucleons,
$\eta^{-1} = n_\gamma/n_B \sim 10^{10}$, deuterium production is delayed past
the point where the temperature has fallen below the deuterium binding energy,
$E_B = 2.2$ MeV (the average photon energy in a blackbody is
${\bar E}_\gamma \simeq 2.7 T$).
When the quantity $\eta^{-1} {\rm exp}(-E_B/T) \sim 1$ the rate for
deuterium destruction (${\D + \gamma \rightarrow p + n}$)
finally falls below the deuterium production rate and
the nuclear chain begins at a temperature $T \sim 0.1 MeV$.

The dominant product of big bang nucleosynthesis is \he4 resulting in an
abundance of close to 25\% by mass. This quantity is easily estimated by
counting the number of neutrons present when nucleosynthesis begins.
When the weak interaction rates freeze-out, at $T \approx 0.8$ MeV,
the neutron to proton ratio is about 1/6. When free neutron decays
are taken into account prior deuterium formation, the ratio drops to
$(n/p) \approx 1/7$. Then simple counting yields a \he4  mass fraction
\be
Y_p = {2(n/p) \over \left[ 1 + (n/p) \right]} \approx 0.25
\label{ynp}
\ee

In the standard model,
 the \he4 mass fraction
depends primarily on the baryon to photon ratio,
$\eta$ as it is this quantity which determines the onset of nucleosynthesis
via deuterium production. But because the $(n/p)$ ratio is only
weakly dependent on $\eta$, the \he4 mass fraction is relatively
flat as a function of $\eta$. The change due
to the uncertainty in the neutron half-life is small
(this effect is shown in Fig. 1). When we go beyond the standard model, the
\he4 abundance is very sensitive to changes in the expansion rate which
can be related to the effective number of neutrino flavors as will
be discussed below. Lesser amounts of the other light elements are produced:
D and \he3 at the level of about $10^{-5}$ by number, and \li7 at the level of
$10^{-10}$ by number.

\begin{figure}[htbp]
\vspace*{13pt}
\centerline{\vbox{\hrule width 5cm height0.001pt}}
\vspace*{6.8truein}		
\centerline{\vbox{\hrule width 5cm height0.001pt}}
\vspace*{13pt}
\fcaption{{The light element abundances from big bang
nucleosynthesis.}}
\end{figure}

The resulting abundances of the light elements are shown in Figure 1 from the
calculations in ref. 2. The curves for the \he4 mass fraction, $Y$, bracket the
computed range based on the uncertainty of the neutron mean-life which
has been taken as\cite{rpp} $\tau_n = 887 \pm 2$ s.
The \he4 curves have been adjusted
according to the corrections in ref. 4. Uncertainties in the produced \li7
abundances have been adopted from the results in ref. 5. Uncertainties in D and
\he3 production are negligible on the scale of this figure.
The  boxes correspond
to the observed abundances and will be discussed below.  It is clear that
as the observational boxes line up on top of each other, there is an
overall agreement between theory and observations in the range
$\eta_{10} = 10^{10} \eta  = 2.8$ -- 4.5.

\section{The Observations}
\subsection{\he4}
Because helium is produced in stars along with heavier elements,
it is necessary to look for
primordial helium in regions where the stellar processing is minimal, i.e., in
regions where the abundances of elements such as carbon, nitrogen and oxygen
are
very low.  The \he4 abundance in very low metallicity regions is best
determined
from observations in extragalactic HII regions of HeII $\rightarrow$ HeI
recombination lines.  There are extensive compilations of observed
abundances of \he4, N, and O, in many different galaxies \cite{p,evan,iz}. In
Figure 2, the \he4 vs. O/H data  is shown along
with its associated linear fit. This is an updated version of the plot
from ref. 9,  including
data from ref. 8
(details of this fit are given in the last line of Table 1).

\begin{figure}[htbp]
\vspace*{13pt}
\centerline{\vbox{\hrule width 5cm height0.001pt}}
\vspace*{3.3truein}		
\centerline{\vbox{\hrule width 5cm height0.001pt}}
\vspace*{13pt}
\fcaption{{The observed abundances of \he4 vs. O/H in
extragalactic HII regions along with a linear fit to the data.}}
\end{figure}

In Table 1, various fits to the data and subsets of the data are given.
Details concerning the subsets of the data shown (as
well as a more complete discussion on primordial \he4)
can be found in ref. 9. As one
can see there is a considerable degree of stability in these fits, leading to a
2 $\sigma$ upper limit of 0.238 --0.240 for the primordial
abundance of \he4. There is
in addition an overall systematic uncertainty of about 0.005 in $Y_p$.

\begin{table}\begin{center}
\tcaption{Linear Fits for $Y$ vs. $O/H$.}\label{tab:smtab1}

\begin{tabular}{|c|c|c|c|c|c|c|} \hline\hline
Set &  \# Regions & $r$ & ${\chi}^2/dof$ &  $Y_P$  &  $10^{-2}
 \times$ slope &$Y_P^{2\sigma}$ \\ \hline

``All" & 49 & 0.56 & 0.78 & $.234 \pm .003$ & $ 1.14 \pm 0.24$ & 0.239 \\
1st cut & 41 & 0.51 & 0.61 & $.232 \pm .003$ & $ 1.38 \pm 0.36$& 0.238 \\
-outliers & 34 & 0.45 & 0.70 &$.232 \pm .003$ & $1.39 \pm 0.38$& 0.238 \\
2nd cut & 21 & 0.41 & 0.64 & $.229 \pm .005$ & $2.37 \pm 1.13$ & 0.238 \\
-outliers & 19 & 0.40 & 0.70 & $.229 \pm .005$ & $2.42 \pm 1.15$ & 0.238\\
C & 22 & 0.35 & 0.71 & $.232 \pm .003$ & $1.58 \pm 0.54$ & 0.238 \\
IZ + ``All"& 55 & 0.53 & 0.82 &  $.236 \pm .002$ & $1.02 \pm 0.23$& 0.240 \\
IZ + 1st cut & 47 & 0.49 & 0.69 & $.234 \pm .003$ & $1.24 \pm 0.33$ & 0.239 \\
\hline\hline
\end{tabular}
\end{center}
\end{table}

The size of the assumed systematic errors has recently been
questioned\cite{cst,sg}. Several sources for the
systematic errors have been discussed and it has been argued that the
systematic error is infact significantly larger than 0.005. Though
the cumulative effect from several systematic uncertainties is
unclear, linearly adding the various systematic
errors assumes that they are correlated and will surely result in an
over-estimate of  the true
error. On the other hand,
the effects of detailed radiative transfer may be more important
than previously thought\cite{sg} in determining the \he4 abundance,
 and pending new results concerning these
calculations it is premature to estimate the size of the correction
to $Y$, if any. For now, we will continue to assume that
$\sigma_{\rm systematic} = 0.005$ though we must bear in mind that
the true error may be somewhat larger.

As was noted, from Table 1 it appears that the results for $Y_p$, the
intercept of the linear fit to the $Y$ vs. $O/H$ data, are quite stable.
In order to test the stability of the fit\footnote{Work done
in collaboration with G. Steigman},~we have randomly
sampled 49 points from the 49 observed points from refs. 6 and 7
and included in fig. 2.
10,000 separate sets of 49 points were obtained.
We then performed a least squares linear fit to each set of points.
The helium fraction for the zero metallicity point for each of the fits
is illustrated in fig. 3.  The fits are strongly peaked
around the mean value of $.232\pm .003$ (.005 at 95\% CL) in
remarkable agreement with the simple linear fit in the first line of Table 1.
We can conclude that there are apparently no observed points that are
strongly influencing the fit. We have similarly run tests on the
first cut data set of 41 points, with the resulting mean
value for the intercepts of $0.232 \pm 0.003$ (0.007 at 95\% CL).

\begin{figure}[t]
\vspace*{13pt}
\centerline{\vbox{\hrule width 5cm height0.001pt}}
\vspace*{4.3truein}		
\centerline{\vbox{\hrule width 5cm height0.001pt}}
\vspace*{13pt}
\fcaption{{The distribution of intercepts from $10^4$ linear fits
of $Y$ vs. O/H based on a random selection of data from the full data set of
49 extragalactic HII regions used in ref. 9.}}
\end{figure}

Subsequent to the analysis in ref. 9, concerning primordial \he4,
there has been some newer data\cite{iz} on \he4 in low metallicity
extragalactic HII regions.  Here we show how these results
modify the above fits of \he4 vs O/H. The data in Izatov et al.\cite{iz}
consists of \he4, O/H, and N/H measurements in 10 extragalactic
HII regions. Four of these regions had been observed in
Skillman et al. \cite{evan} and these have been averaged together with the
new data.  Thus there are a total of 6 new points in the data set.
We will not enter into the discussion regarding the methods of
data reductions used by Izatov.  Here we have used their data based
on having taken electron densities from the ratio of collisionally
excited SII lines
(see ref. 8, for details).  Though there is a slight shift upwards, the shift
is entirely consistent within the statistical uncertainties.
One could however make a case that based on this set of data
(the combined set) that the 2 $\sigma$ upper limit to $Y_p$ is 0.239
which moves up to 0.244 when a systematic uncertainty of 0.005 is added.

\begin{figure}[t]
\vspace*{13pt}
\centerline{\vbox{\hrule width 5cm height0.001pt}}
\vspace*{3.3truein}		
\centerline{\vbox{\hrule width 5cm height0.001pt}}
\vspace*{13pt}
\fcaption{{The observed abundances of \he4 vs. N/H in
extragalactic HII regions along with a linear fit to the data.}}
\end{figure}

\begin{table}[t]\begin{center}
\tcaption{Linear Fits for $Y$ vs. $N/H$.}\label{tab:smtab2}

\begin{tabular}{|c|c|c|c|c|c|c|} \hline\hline
Set &  \# Regions & $r$ & ${\chi}^2/dof$ &  $Y_P$  &  $10^{-2}
 \times$ slope &$Y_P^{2\sigma}$ \\ \hline
All & 49 & 0.66 & 0.66 & $.236 \pm .002$ & $ 1.72 \pm 0.33$ & 0.240\\
1st cut & 41 & 0.57 & 0.58 & $.234 \pm .002$ & $ 2.71 \pm 0.68$  & 0.239 \\
-outliers & 34 & 0.48 & 0.69 &$.234 \pm .003$ & $2.77 \pm 0.76$ & 0.239\\
2nd cut & 21 & 0.47 & 0.63 & $.231 \pm .004$ & $4.85 \pm 2.27$ &0.239 \\
-outliers & 19 & 0.44 & 0.70 & $.232 \pm .004$ & $4.79 \pm 2.29$ & 0.239 \\
C & 22 & 0.46 & 0.60 & $.233 \pm .003$ & $3.62 \pm 1.17$ & 0.238 \\
IZ + ``All"& 55 & 0.61 & 0.74 &  $.239 \pm .002$ & $1.51 \pm 0.31$& 0.242 \\
IZ + 1st cut & 47 & 0.52 & 0.70 & $.237 \pm .002$ & $2.35 \pm 0.64$ & 0.241 \\
\hline\hline
\end{tabular}
\end{center}
\end{table}

It is also of interest to compare the correlation of \he4 with
N/H. Indeed, a comparison of N/H vs O/H is also of interest\cite{OSt,don}.
Since the stars which produce oxygen do not coincide exactly with those
which produce nitrogen, the linear regression of $Y$ vs N/H is another
consistency check.  For the same data as discussed above, the corresponding
results to $Y$ vs N/H are given in Table 2. Again, we see that there is
a considerable degree of consistency within the data. The 2 $\sigma$ upper
limit
is slightly higher. The stability of the $Y$ vs. N/H fits was also checked
( by the same method used to generate Fig.3 ). For the full data set
of 49 points from ref. 9 (line 1 of Table 2), the mean
value of intercepts found was
0.234 $\pm 0.002$ (0.005 at 95\% CL). For the first cut set of points
(line 2 of Table 2), the resulting mean value was also 0.234 with a
slightly larger uncertainty $\pm 0.003$ (0.006 at 95\%CL).

 Looking at Tables 1 and 2, it is not unreasonable
to take as the best estimate to the primordial mass fraction of \he4
\be
Y_p = 0.234 \pm 0.003 \pm 0.005
\label{YP}
\ee
with a 2 $\sigma$ upper limit of 0.240 and 0.245 when systematic uncertainties
are included. Thus we will assume a
range (2 $\sigma$ plus systematic) of 0.223 -- 0.245 for $Y_p$ which is shown
in
Fig. 1 as the large box bracketing the \he4 curves.

Before concluding the discussion on \he4, we would like to update
an older analysis\cite{wssok} concerning the correlation of \he4 with C/H.
It was suggested\cite{sgs} that because carbon (and nitrogen) are
produced in intermediate mass stars, that the correlation between
these elements and \he4 might be more significant than that of \he4 and O/H.
The correlation between \he4 and C/H was tested in Walker et al.\cite{wssok}
where the fit to $Y$ vs. C/H gave similar results as for $Y$ vs O/H and N/H
albeit with larger uncertainties to due to the poor statistics (there
were only 6 extragalactic  HII regions with measured C/H).  Recently
there have been new measurements\cite{GS} of C/O ratios in seven regions.
The data are somewhat better than previous data (at least for small C/H)
and are shown in Fig. 5.  Once again, $Y$ vs C/H is of limited value because of
the paucity of data, yet yields a consistent (with O/H and N/H) fit
with an intercept $Y_p = 0.231 \pm 0.004$ with a 2 $\sigma$ upper limit of
0.240; the same as we have concluded above from O/H and N/H.

\begin{figure}[h]
\vspace*{13pt}
\centerline{\vbox{\hrule width 5cm height0.001pt}}
\vspace*{3.1truein}		
\centerline{\vbox{\hrule width 5cm height0.001pt}}
\vspace*{13pt}
\fcaption{{The observed abundances of \he4 vs. C/H in
extragalactic HII regions along with a linear fit to the data.}}
\end{figure}

\subsection{D and \he3}
It is more difficult to
compare the primordial deuterium  and \he3 abundances
with the observations. Despite the fact that
all observed deuterium is primordial,  deuterium is destroyed in stars.
A comparison between the predictions
of the standard model and observed solar and interstellar values of deuterium
must be made in conjunction with models of galactic chemical evolution.
The problem concerning \he3 is even more difficult.  Not only
is primordial \he3 destroyed in stars but it is very likely that low mass stars
are net producers of \he3. Thus the comparison between theory and observations
is
complicated not only by our lack of understanding regarding chemical evolution
but also by the uncertainties of the production of \he3  in stars.
Furthermore, due to the large uncertainty in the age of the galaxy, it
is difficult to directly compare the observed and modeled values of
deuterium and \he3 since the exact galactic age at which the
pre-solar epoch occurs is unknown.  We will return to this problem below.

It appears that D/H has decreased
over the age of the galaxy.  The (pre)-solar system abundance of deuterium
is inferred from a variety of \he3 measurements in the solar wind,
lunar soil samples and meteoritic samples.  There are two components of
\he3: one associated with the pre-solar abundance of
deuterium and \he3 (the deuterium was converted in the early
premain-sequence stage of the sun); while the second is \he3 trapped
in meteorites which represents the true solar value of \he3
\be
\left({{\rm D + \he3} \over H}\right)_\odot  =  (4.1 \pm 1.0) \times 10^{-5}
\ee
\be
\left({{\he3} \over H}\right)_\odot  =  (1.5 \pm 0.3) \times 10^{-5}
\ee
The difference between the two \he3 measurements represents the
presolar D abundance and is\cite{geiss} D/H $\approx (2.6 \pm 1.0)
\times 10^{-5}$.  On the other hand,
the present ISM
abundance of D/H is\cite{linsky}
\be
{\rm D/H} = 1.60 \pm 0.09 {}^{+.05}_{-.10} \times 10^{-5}
\ee
It
is this lower limit on D/H (since D is only destroyed) that provides us
with the the upper bound on $\eta$.  It is shown as the lower
side of the D and \he3 box in Fig. 1. Thus, if $\eta_{10}$
is in the range 2.8 -- 3.9 then the primordial abundance of D/H is between 4.5
-- 8 $\times 10^{-5}$, and it would appear that significant destruction of
deuterium is necessary.

 Note that  there are reported detections of D in  high redshift,
low metallicity quasar absorption systems\cite{quas1,quas2}.
In principle, these measured abundances should represent the primordial one.
Unfortunately, at this time, the two observations do not yield a single
value for D$_p$.  In one\cite{quas1}, D/H $\approx 1.9 - 2.5 \times 10^{-4}$
while the second gives\cite{quas2}, D/H $\approx 1 - 2 \times 10^{-5}$.
(One should note that it has been argued that these types of measurements
are subject to large systematic corrections\cite{lt}, thus it may not be
all that surprising that the two measurements differ.)
The former observation is shown in Fig. 1 by the small box on
the D/H curve at a
value of $\eta_{10} \approx 1.5$.  As one can see the corresponding value of
$Y_p$ (at the same value of $\eta$) is in excellent agreement with the data.
\li7 is also acceptable at this value as well. Due to the still some what
preliminary status of this observation it is premature
to fix the primordial abundance with this value. A high value
for the D abundance would require an even greater degree of D destruction over
the age of the galaxy. The lower measurement for D/H is problematic for
both \he4 and \li7.

To trace the evolution of D and \he3, we can make use of some simple
models of galactic chemical evolution.
For a general review on chemical evolution see ref. 20.  We
shall adapt her same basic notation in the following discussion.
 One can relate the deuterium abundance
at any time to the gas mass fraction noting that deuterium is only
destroyed in stars.
The equation for the gas mass in the galactic disk is
\begin{equation}
{{dM_G}\over{dt}} = e(t) -\psi(t) + f(t). \label{gas}
\end{equation}
In this equation, $e(t)$ is the gas ejected from stars at the end of
their lifetimes either in planetary nebulae or supernova events,
\begin{equation}
e(t) =
\int\limits_{m(t)}^{m_{upp}}\!{(m-m_R)\psi(t-\tau_m)\phi(m)}\,dm. \label{eject}
\end{equation}
In these equations,
$f(t)$ is the infall rate onto the disk of gas whose origins is
external to the galaxy.
$\phi(m)$ is the initial mass function (IMF), and
$\psi(t)$
is the star formation rate (SFR).  $m_{upp}$ is the upper mass limit on
$\phi(m)$.
$\tau_m$ is the lifetime of a star of mass $m$,
and $m_R$ is the remnant mass left over by a star of mass $m$.
$m(t)$ is is the
mass
of a star which at time $t$ is returning gas back into the ISM.

 In order to model the deuterium evolution, we
consider $D$ to be totally astrated within
stars.
We may extend equation (\ref{gas}),
\begin{equation}
{d{\rm D}\over {dt}} = -e(t){\rm D} + f(t)({\rm D}_f-{\rm D}). \label{deut}
\end{equation}
where D$_f$ is the mass fraction of D in the infall gas.
Using the instantaneous recycling approximation (an assumption which
ignores delays due to finite stellar lifetimes, and is an acceptable
approximation at early times and for elements which are produced in
massive stars whose lifetimes are short) we can combine Eqs. (\ref{gas})
and (\ref{deut}) to give the total amount of deuterium destruction
relative to its initial primordial value D$_p$\cite{ot}
\be
{\rm {D \over ~D_p}} = \sigma^{R/(1-R)}
\ee
where $\sigma$ is the gas mass fraction and the return fraction $R$ is given by
\be
R = \int\limits_{m(t)}^{m_{upp}}\!{(m-m_R)\phi(m)}\,dm
\ee
By a suitable choice for the IMF (and SFR) a considerable amount of
deuterium may have been destroyed\cite{ddest,vop}.

As we noted above, to match the abundances in a model of galactic
evolution to the observations requires knowledge of the time at which the
solar system was formed.
In order to solve this problem, we\cite{sco} have incorporated the production
of the long-lived r-process nuclear chronometers $^{232}Th$, $^{235}U$, and
$^{238}U$ into chemical evolution models.  Previously, these
chronometers have been used in simple chemical evolution models in an
attempt to constrain the over-all galactic age\cite{ctt}.
This is accomplished by finding a time at which the modeled ratios
$^{232}Th/^{238}U$ and $^{235}U/^{238}U$ match (if ever) the values of these
ratios determined from meteorite data ($^{232}Th/^{238}U$ = 2.32
and $^{235}U/^{238}U$ = .317\cite{ae}).
This time is taken to be the pre-solar
epoch, and 4.6 Gyrs. later, today.

A full description of how nuclear chronometers may be incorporated in
general into galactic chemical evolution models can be found in
ref. 25.  We summarize the main points below.
Nuclear chronometers are incorporated into the calculation by an
extension of the basic equation of the gas mass in the galactic disk
discussed above.
We may extend equation (\ref{gas}) to determine the rate of change in the
number of nuclear species $A$,
\begin{equation}
{{dN_A}\over{dt}} = P_A\psi(t) - {{\psi(t)N_A}\over M_G} +
{{e(t)N_A}\over
M_G} + {f\over M_G}{Z_f\over Z}N_A - \lambda_A N_A. \label{chron}
\end{equation}
In this equation, $P_A$ is the number of newly synthesized nuclei of
species $A$
per unit mass going into star formation.  The relative production
ratios,
$P_{232}/P_{238}$, and $P_{235}/P_{238}$, have been determined by
supernova models\cite{ctt},
$P_{232}/P_{238}$
= 1.60, and $P_{235}/P_{238}$ = 1.16. Since we have assumed the infall
gas to be primordial, $Z_f$, the metallicity of the infall
gas, has been taken to be zero. And finally,
$\lambda_A$ is the rate of decay
of nuclear species $A$.  As a simplification, the amount of nuclear
species ``A" which decays while locked up in stars was not taken into
account equation (3).  Only large stars will contribute to the
production of these nuclear species and they have lifetimes which are
very short when compared with the decay times.

Equations (\ref{gas}) and (\ref{chron}) can then be solved numerically.
 In order to
satisfy observational constraints, the value of $N_A(0)$ is allowed to be
non-zero.  This corresponds to an initial enrichment ($S_0$) in the galactic
disk of metals.  This solves the so-called G-dwarf problem; where
observationally, there appear to be very few metal poor dwarf stars in
the disk\cite{lb}.

We have found a wide range of
chemical evolution models
capable of destroying $D$ by a factor of 5, which also satisfy our
nuclear chronometer constraint.  These models result in a
range of galactic ages ($11.1-15.0$ Gyrs). The result of one of these
models is
shown in Fig. 6.

\begin{figure}[htbp]
\vspace*{13pt}
\centerline{\vbox{\hrule width 5cm height0.001pt}}
\vspace*{3.1truein}		
\centerline{\vbox{\hrule width 5cm height0.001pt}}
\vspace*{13pt}
\fcaption{{The evolution of D/H in several models of galactic chemical
evolution\cite{sco}.}}
\end{figure}

There are however potential problems for \he3. A lower limit on $\eta$ was
derived\cite{ytsso} by noting that although stars can destroy \he3, even
very massive stars, still preserve at least 25 \% of the initial D + \he3.
(It is the sum of D and \he3 that is important as D is burned to \he3 in the
premain-sequence phase of stars.)  A value of $\eta_{10}$ lower than 2.8 would
yield (D+\he3)/H $> 10^{-4}$ so that even if the maximal amount of \he3
is destroyed, it would still exceed the presolar value\cite{geiss} of
(D+\he3)/H
$\approx 4.1 \pm 1.0 \times 10^{-5}$. But, in low mass stars, \he3 is produced
rather than destroyed.  A calculation\cite{it} of the final \he3 abundance
relative to the initial abundance of D + \he3 gives,
\begin{equation}
(^3{\rm He/H})_f = 1.8 \times 10^{-4}\left({M_\odot \over M}\right)^2
+ 0.7\left[({\rm D+~^3He)/H}\right]_i \label{hel3}
\end{equation}
so that a 1 M$_\odot$ star produces 2.7 times as much \he3 as the initial D +
\he3 (for (D + \he3)$_{\rm initial} = 9 \times 10^{-5}$). This would lead to an
evolutionary behavior\cite{orstv} of the type shown in Fig. 7. The chemical
evolution model has been chosen so that
 D/H agrees with the data and assumes that $\eta_{10} = 3$.
The problem being emphasized concerns \he3 and can be seen by comparing
the solid curve with the filled diamonds.

A number of questions regarding the \he3 discrepancy can be raised. First, one
can ask whether or not the formula for
\he3 production is valid.  There is indeed evidence that \he3 is produced in
planetary nebulae\cite{rbw} where abundances are found as high as \he3/H $\sim
10^{-3}$. Secondly, one may ask how uniform are the \he3 measurements. As it
turns out, they are in fact not very uniform\cite{bbbrw} and show variations as
large as a factor of 5 between different HII regions.  Indeed, there may even
be a correlation between the size (mass) of the region and the amount of \he3
observed\cite{orstv}. Thus it may be possible that the solar values are
depleted in \he3. It is hoped that future \he3 observations will help to
resolve this puzzle.

\begin{figure}[t]
\vspace*{13pt}
\centerline{\vbox{\hrule width 5cm height0.001pt}}
\vspace*{3.1truein}		
\centerline{\vbox{\hrule width 5cm height0.001pt}}
\vspace*{13pt}
\fcaption{{The evolution of
D/H (dashed curve), \he3/H (solid curve) and (D + \he3)/H (dotted curve)
  as a function of time.  Also shown are the data at the solar epoch
$t \approx 9.6$ Gyr and today for
D/H (open squares), \he3/H (filled diamonds) and
 (D + \he3)/H (open circle).\cite{orstv}.}}
\end{figure}

Before we conclude with D and \he3, we address the overall issue regarding
consistency of the data and the big bang nucleosynthesis calculations.
Recently, it has been suggested\cite{hata2} that there is a crises in
 big bang nucleosynthesis because of an underlying inconsistency between the
``predicted" value of $\eta$ (from  deuterium) and the \he4 observations.
The conclusion was that to a high degree of confidence, the standard
model with three light neutrino flavors is excluded.  Though we will
return to neutrino limits in the next section, we can here dispell the notion
of
a major inconsistency. The crises in ref. 33, arises from using a
pedagogical expression\cite{ped,ytsso}
 relating deuterium and \he3 at a given time
assuming that a fraction $f$ of the total gas mass passed through stars
at time $t$,
and that a fraction $g_3$ of the initial D + \he3 survives as \he3
when \he3 production is neglected
\be
{\rm \left({D + \he3 \over H} \right)_p \le \left({D \over H} \right)_t}
+ {1 \over g_3}{\rm  \left({\he3 \over H} \right)_t} - \left({1-g_3
\over g_3} \right)f{\rm  \left({\he3 \over H} \right)_p}
\label{ex}
\ee
This expression has been used\cite{ytsso} extensively to place a
lower bound on $\eta$ by inserting the observed values for
D/H and \he3/H at the solar epoch.  This is what leads to $\eta_{10}
> 2.8$ when $g_3 > 0.25$
and corresponds to the upper limit on the D + \he3 box in
Fig. 1.  To be sure, this expression indicates that the best fit\cite{hat}
to deuterium is a low value for D/H ($\approx 2.3 \times 10^{-5}$).
Clearly for the corresponding value of $\eta$, there is a significant
discrepancy with \he4.  However, rather than to conclude on this basis
that there is a problem with the standard model, or that we have learned
something about neutrino physics, it is evident to us that
that what we have learned is that there is something wrong
with the expression (\ref{ex}). Indeed it is known that even simple models
of galactic evolution with $g_3 \sim 0.3$ (i.e. neglecting \he3 production)
can give a good match to the D and \he3 solar data\cite{vop}.  In this case,
the discrepancy between the prediction and observations of \he4 is at most
at the 2 $\sigma$ level.  Of course when \he3 production is included, the
problem is more acute\cite{orstv} requiring, perhaps, more dramatic changes
in the chemical evolution models of the galaxy\cite{cosstv}.

\subsection{\li7}

Finally we turn to \li7.  Over the last several years, there has been a
considerable increase in the number of \li7 observations \cite{li}.
 \li7 in old,
hot, population II stars, is found to have a very nearly  uniform abundance.
For
stars with a surface temperature $T > 5500$ K and a metallicity less than about
1/20th solar, the  \li7 abundance shows little or no dispersion beyond what is
consistent with errors of individual measurements.
 A large quantity of data has been obtained from
a variety of sources\cite{li,nrs} and consists of observations of
55 different halo stars. The corresponding mean \li7 abundance is [Li] = 12
+ $\log$ Li/H = 2.08 $\pm$ 0.02 or
the abundance by number is Li/H =
($1.22
\pm .04)
\times 10^{-10}$. Systematic errors however dominate the uncertainty in the
\li7
abundance.

Furthermore there has been considerable attention  to trends in the data.
Namely, it appears that there may be real correlations in the \li7
abundances with respect to temperature and metallicity\cite{nrs,thor}.
It has been argued that to avoid any temperature dependence which may
mask a real metallicity dependence, one should renormalize
the derived \li7 abundances to a single temperature, say 6200K.
For the data discussed above, there is only a weak dependence on temperature,
[Li] = 1.20 $\pm$ 0.44 + (14 $\pm 7) \times 10^{-5} T$.
This fit is a modest improvement (at the 87 \% CL) over the straight
average. ``Correcting" the data by sliding each point up to 6200K,
along the line given by the fit
above, one finds on the average a slightly higher \li7 abundance,
[Li] = 2.13 $\pm$ 0.02.  If one were to take the raw data, (ie.
\li7 abundances uncorrected for differing temperatures) then only a
questionable correlation with metallicity would result
[Li] = 2.14 $\pm$ 0.05 + (0.025 $\pm$ 0.023)[Fe/H], where
[Fe/H] is the log of the iron abundance relative to the solar iron
abundance.
In contrast, when the corrected lithium abundances are used a
more significant correlation appears, [Li] = 2.24 $\pm$ 0.05 +
(0.051 $\pm$ 0.022)[Fe/H]. This fit represents an improvement at the 98\%
CL over the straight average.

The derived \li7 abundance is quite sensitive to the assumed
temperatures of the stars. Most authors have obtained abundances
through color-based temperatures from photometry.  In addition, the results
depend on the particular stellar model used.  In Thorburn's\cite{thor}
analysis of over 70 halo stars, she finds a systematically higher temperature
and hence \li7 abundance.  In this sample [Li] = 2.26 $\pm 0.01$ or
Li/H =
($1.82
\pm .04)
\times 10^{-10}$.

\begin{figure}[t]
\vspace*{13pt}
\centerline{\vbox{\hrule width 5cm height0.001pt}}
\vspace*{3.3truein}		
\centerline{\vbox{\hrule width 5cm height0.001pt}}
\vspace*{13pt}
\fcaption{{\li7 data shown as a function of temperature. [Li] =
$12 + \log Li/H$. The horizontal line is the mean of this data.}}
\end{figure}

The same correlation (and to higher degree) between Li and temperature and
metallicity is present in the Thorburn data.
The correlation with temperature in that data appears to be very strong
[Li] = 0.06 $\pm$ 0.37 + (36 $\pm 6)\times 10^{-5} T$
which leads to a higher mean \li7 abundance [Li] = 2.33 $\pm$ 0.01.
There is also marked effect on the fit with respect to [Fe/H].
Using the raw \li7 abundances,
[Li] = 2.49 $\pm$ 0.09 + (0.087 $\pm$ 0.036)[Fe/H]
and using the corrected \li7 abundances gives
[Li] = 2.73 $\pm$ 0.08 + (0.16 $\pm$ 0.03)[Fe/H].  These fits all have
significantly lower $\chi^2$'s than does the straight average.
Of course these trends make it more difficult to determine with any
certainty the true primordial abundance.

Recently however, the method of obtaining temperatures by means of
broad-band photometry has been criticized\cite{fag} and a strong
case has been made to use a spectroscopic method based on Balmer
line profiles.  This method was employed\cite{mol} in reanalyzing a relatively
large set of halo stars (24 with T $>$ 5700K with [Fe/H] $<$ -1.4).
In this data set the mean value for \li7 is [Li] = 2.21 $\pm$ 0.02
or Li/H = (1.62 $\pm 0.07) \times 10^{-10}$. However in this case,
there is absolutely no evidence for a trend with respect to temperature,
[Li] = 1.90 $\pm$ 0.57 + (5 $\pm 9) \times 10^{-5} T$ which corrects the mean
value to [Li] = 2.22 $\pm$ 0.02. More importantly there is no correlation
with metallicity either.  The uncorrected data gives
[Li] = 2.20 $\pm$ 0.10 + (0.00 $\pm$ 0.04)[Fe/H] while the ``corrected" data
gives
[Li] = 2.23 $\pm$ 0.10 + (0.01 $\pm$ 0.04)[Fe/H].

Though it is difficult to place a specific value on the total systematic
error in any of the \li7 abundances, from the spread in the obtained
values it would appear that a fair estimate of the systematic uncertainty
is about 0.1 dex. It may indeed be smaller in the last analyses\cite{mol}.
Taken together with what should be considered the best estimate for the
\li7 abundance in halo dwarfs, we arrive at a final \li7 abundance of
\be
{\rm [Li] = 2.21 \pm 0.02 \pm 0.1}
\ee
 corresponding to a range for  Li/H
between $\sim 1.2 ~{\rm and}~ 2.2 \times 10^{-10}$ which is shown in Fig. 1
as the observed range for \li7. These values (and
their uniformity) should be compared with observations of Li in younger stars
where the abundance can be much larger (by an order of magnitude) and shows
considerable dispersion. Two key questions remain however: how much of
the observed Li is primordial (since Li is known to be produced); and how much
of the primordial Li remains in the stars where Li is observed?

Aside from the big bang, Li is produced together with Be and B in cosmic ray
spallation of C,N,O by protons and $\alpha$-particles.  Li is also produced by
 $\alpha-\alpha$ fusion.  Be and B have recently been observed in these same
pop II stars and in particular there are a dozen or so, stars in which both
Be and \li7 have been
observed.  Thus Be (and B though there is still a paucity of
data) can be used as a consistency check on primordial Li \cite{check}. Based
on
the Be abundance found in these stars, one can conclude that no more than 10-20
\% of the \li7 is due to cosmic ray nucleosynthesis leaving the remainder
(an abundance near $10^{-10}$) as primordial. It is also possible however, that
some of the initial Li in these stars has been depleted.  Standard stellar
models\cite{del} predict that any depletion of \li7 would be accompanied by a
very severe depletion of \li6.  Until recently, \li6 had never been observed in
hot pop II stars. The observation\cite{li6o} of \li6 (which turns out to be
consistent with its origin in cosmic-ray nucleosynthesis and with a small
amount
of depletion as expected from standard stellar models) is  a good indication
that
\li7 has not been destroyed in these stars\cite{li6}.

\subsection{Summary}

Consistency of the standard model of big bang nucleosynthesis relies on the
concordance between theory and observation of the light element abundances for
a single value of $\eta$. We now summarize the constraints on $\eta$ from each
of the light elements.   From the \he4 mass fraction, $Y < 0.240 (0.245)$,
we have that $\eta_{10} < 2.9 (4.5)$ as a $2 \sigma$ upper limit (the higher
value takes into account possible systematic errors). It is appropriate to
note here the sensitivity of the bound on $\eta$ to the assumed \he4
upper limit. When the data of Izatov et al.\cite{iz} are not included,
our upper limit to $Y$ was\cite{OSt} $Y < 0.238 (0.243)$. In this case
the corresponding upper limit to $\eta$ is $\eta_{10} < 2.5 (3.9)$.
Because of the
sensitivity to the assumed upper limit on $Y_p$, the upper limit on $\eta$
from D/H, though weaker is still of value.  From D/H $> 1.5 \times 10^{-5}$
we have $\eta_{10} < 7$.  The lower limit on $\eta$, comes from the upper
limit on D + \he3 and is $\eta_{10} \ga 2.8$ if one ignores \he3 production.
We stress that this limit is only exemplary as the upper limit on
D + \he3 depends critically on models of galactic chemical evolution,
which are far from being understood.
Finally, \li7  allows a broad range for $\eta$ consistent with other light
elements.  When both uncertainties in the reaction rates and systematic
uncertainties in the observed abundances are taken into account, \li7 allows
values of $\eta_{10}$ between 1.3 -- 4.9.
The resulting consistent range for $\eta_{10}$ becomes 2.8 -- 4.5.
These bounds on
$\eta$ constrain the fraction of critical density in baryons,$\Omega_B$
to be
\be
0.01 < \Omega_B < .1
\ee
 for a hubble parameter, $h_o$, between 0.4 -- 1.0. The
corresponding range for
$\Omega_B h_o^2$  is 0.010 -- 0.016.

\section{Constraints on Physic beyond the Standard Model}

Limits on particle physics beyond the standard model are mostly sensitive to
the bounds imposed on the \he4 abundance. As is well known, the $^4$He
abundance
is predominantly determined by the neutron-to-proton ratio just prior to
nucleosynthesis and is easily estimated assuming that all neutrons are
incorporated into $^4$He (see Eq. (\ref{ynp})).
As discussed earlier, the neutron-to-proton
ratio is fixed by its equilibrium value at the freeze-out of
the weak interaction rates at a temperature $T_f \sim 1$ MeV modulo the
occasional free neutron decay.  Furthermore, freeze-out is determined by the
competition between the weak interaction rates and the expansion rate of the
Universe
\begin{equation}
{G_F}^2 {T_f}^5 \sim \Gamma_{\rm wk}(T_f) = H(T_f) \sim \sqrt{G_N N} {T_f}^2
\label{comp} \label{freeze}
\end{equation}
where $N$ counts the total (equivalent) number of relativistic particle
species.
 The presence
of additional neutrino flavors (or any other relativistic species) at
the time of nucleosynthesis increases the overall energy density
of the Universe and hence the expansion rate leading to a larger
value of $T_f$, $(n/p)$, and ultimately $Y_p$.  Because of the
form of Eq. (\ref{comp}) it is clear that just as one can place limits
\cite{ssg} on $N$, any changes in the weak or gravitational coupling constants
can be similarly constrained (for a recent discussion see ref. 46).

In the standard model, the number of particle species entering into eq.
(\ref{comp}) can be written as $N  = 5.5 + {7 \over 4}N_\nu$ (5.5 accounts for
photons and $e^{\pm}$). The observationally derived primordial \he4
abundance from Eq. (\ref{YP}) of
$Y_p = 0.234 \pm 0.003 \pm .005$ translates into a best value for $N_\nu = 2.2
\pm 0.27 \pm .42$ (assuming a central value for $\eta_{10}$ of 3.6,
with $\eta_{10} = 3$, the best value for $N_\nu$ increase to
2.4) which implies a $2 \sigma $ upper limit of 2.74 which is
extended to $N_\nu < 3.16$ when systematics are included.
At face value, such a limit would exclude even a single additional scalar
degree of freedom (which counts as ${4 \over 7}$) such as a majoron unless it
decoupled early enough\cite{oss} so that its temperature, $T_B$ at the time of
nucleosynthesis was suppressed so that $(T_B/T_\nu)^4 < {7 \over 4}(.16) =
.28$.
In models with right-handed interactions, and three right-handed neutrinos, the
constraint is more severe. The right-handed states must have decoupled early
enough to ensure $(T_{\nu_R}/T_{\nu_L})^4 < (.16)/3 \simeq .05$. The
temperature of a decoupled state is easily determined from entropy
conservation, $(T_x/T_\nu) = \left( (43/4)/N(T_d) \right)^{1/3}$. One
additional scalar requires $N(T_d) > 28$ or decoupling must have occurred
above the QCD phase transition at a temperature $T_d > T_{\rm QCD} \sim 200$
MeV.  Three right-handed neutrinos would require $N(T_d) \ga 100$, so that $T_d
> M_W$. If right-handed interactions are mediated by additional gauge
interactions, the associated mass scale becomes $M_{Z'} > O(100)$ TeV!

The limits on $N_\nu$, however,  are sensitive to the upper limit on \he4 which
is in turn sensitive to assumed systematic errors and to the lower bound on
$\eta$.  In addition, the limits described above may be overly
restrictive\cite{ost2}. The best value for $N_\nu$ is 2.2 and may in fact be
unphysical if $\nu_\tau$ is lighter than $\sim 1$ MeV, as is quite likely.  In
this case the limits on $N_\nu$ must be accordingly renormalized as has been
discussed extensively in ref. 48.
In Fig. 9, the effect of renormalizing the limit on $N_\nu$ is shown.
In Fig. 9, the central value of $Y_p = 0.232$ was assumed.  For the
higher value of 0.234, an additional $\approx 0.2$ should be added
to the limits on $N_\nu$.

\begin{figure}[htbp]
\vspace*{13pt}
\centerline{\vbox{\hrule width 5cm height0.001pt}}
\vspace*{3.3truein}		
\centerline{\vbox{\hrule width 5cm height0.001pt}}
\vspace*{13pt}
\fcaption{{The 95 \% CL upper limit on $N_\nu$ as a function of
the systematic uncertainty in $Y_p$. The solid (dashed) curves correspond to
the condition that $N_\nu\ge$ 2 (3). Each of these two cases is shown for three
choices of $\eta_{10}$: 1.5, 2.8, and 4.0. The smaller values of $\eta$
yield weaker upper limits and systematic errors have been
described by a top-hat distribution\cite{ost2}.}}
\end{figure}

\section{Summary}

In summary, we have argued for the overall agreement between theory and
observations as they pertain to the light element abundances as well as
concordance between big bang nucleosynthesis and galactic cosmic-ray
nucleosynthesis.  There is now quite a bit of
 data on \he4 which appears to be coherent.
Because the inferred primordial \he4 abundance found from
extrapolating the data to zero metallicity gives a somewhat low value,
smaller values for the baryon to photon ratio are preferred.  In
contrast, there may be a problem with \he3.  The data in the ISM
shows a large dispersion, which may be an indication of possible local
pollution, and spans a factor of $\sim 5$ in abundance.  Additional
data is clearly needed here before
strong conclusions can be made regarding potential problems.
The solar value of \he3 seems to indicate that little \he3 is
produced over the early history of the galaxy.  While higher values
of the baryon to photon ratio certainly help in this direction,
we believe that this is ultimately a problem to be resolved by
chemical evolution models.  \li7, because of relatively large allowed
range (dominated by systematic uncertainties) allows a wide
range for the baryon to photon ratio.

There are however, open issues: Are the quasar line-of-sight
measurements\cite{quas1,quas2} of D/H real; Why isn't there more \he3,
particularly in
the solar system; Can the statistical and systematic errors in \he4
measurements be reduced; Can the large systematic errors in the \li7 abundance
be reduced?  Clearly new data will be necessary to resolve these problems.
Nevertheless, in spite of these uncertainties, nucleosynthesis is still able to
set strong constraints on physics beyond the standard model.

\nonumsection{Acknowledgments}
\noindent
We would like to thank
J. Audouze, M. Cass\'e, B. Fields, R. Rood, D.N. Schramm, E. Skillman,
G. Steigman, J. Truran, and
E.Vangioni-Flam for many enjoyable and useful conversations.
This work was supported in part by  DOE grant DE-FG02-94ER40823.

\nonumsection{References}

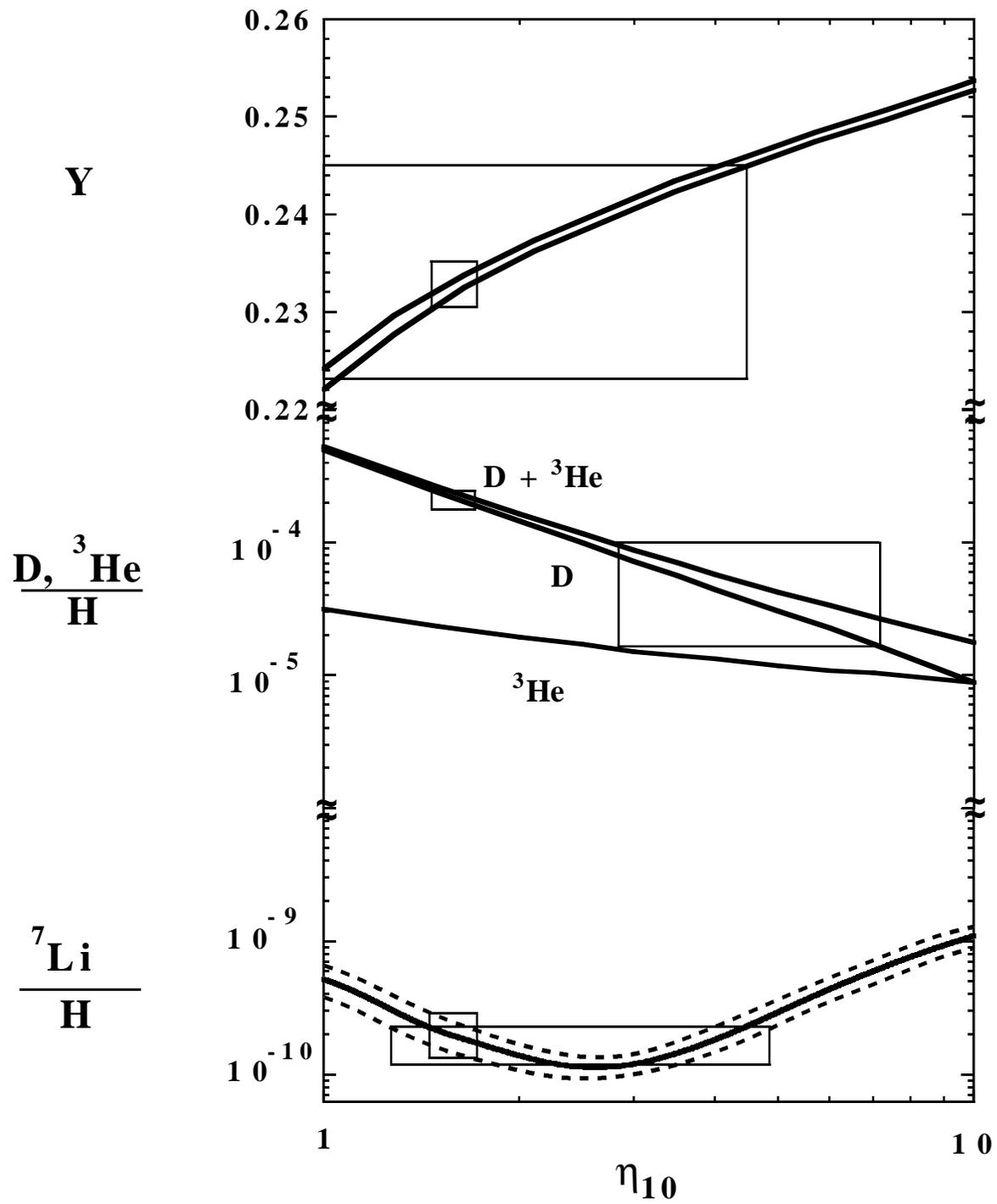
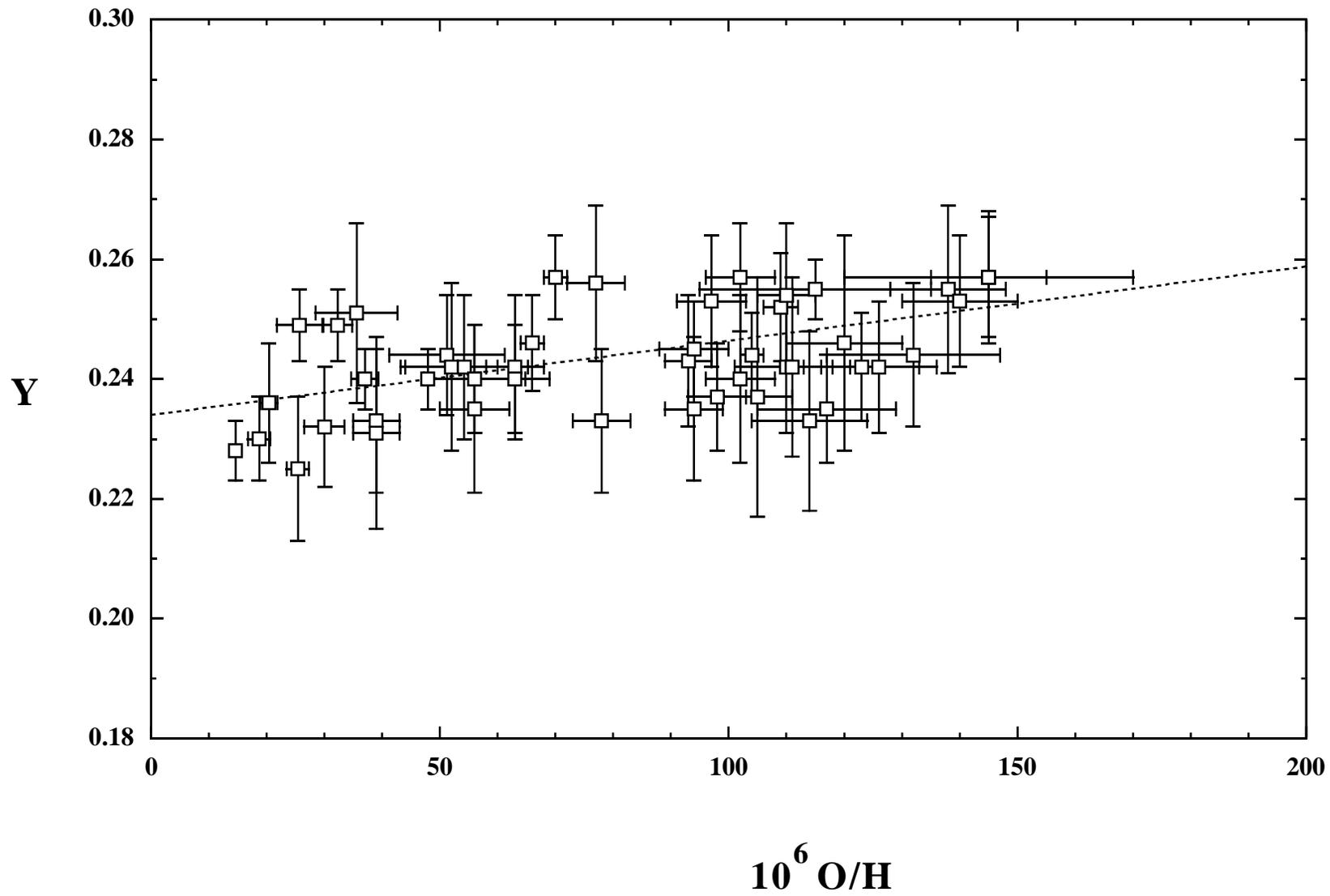
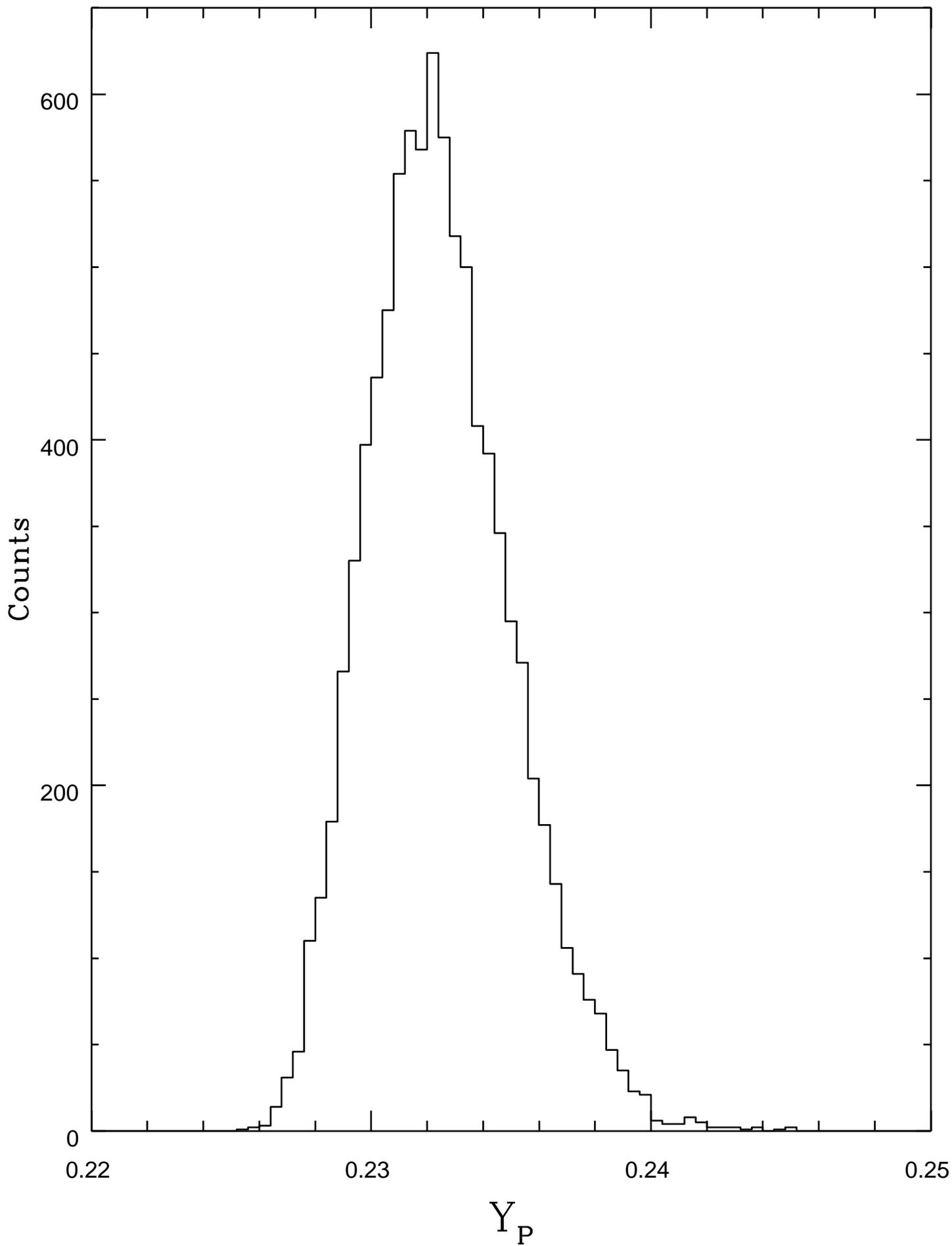
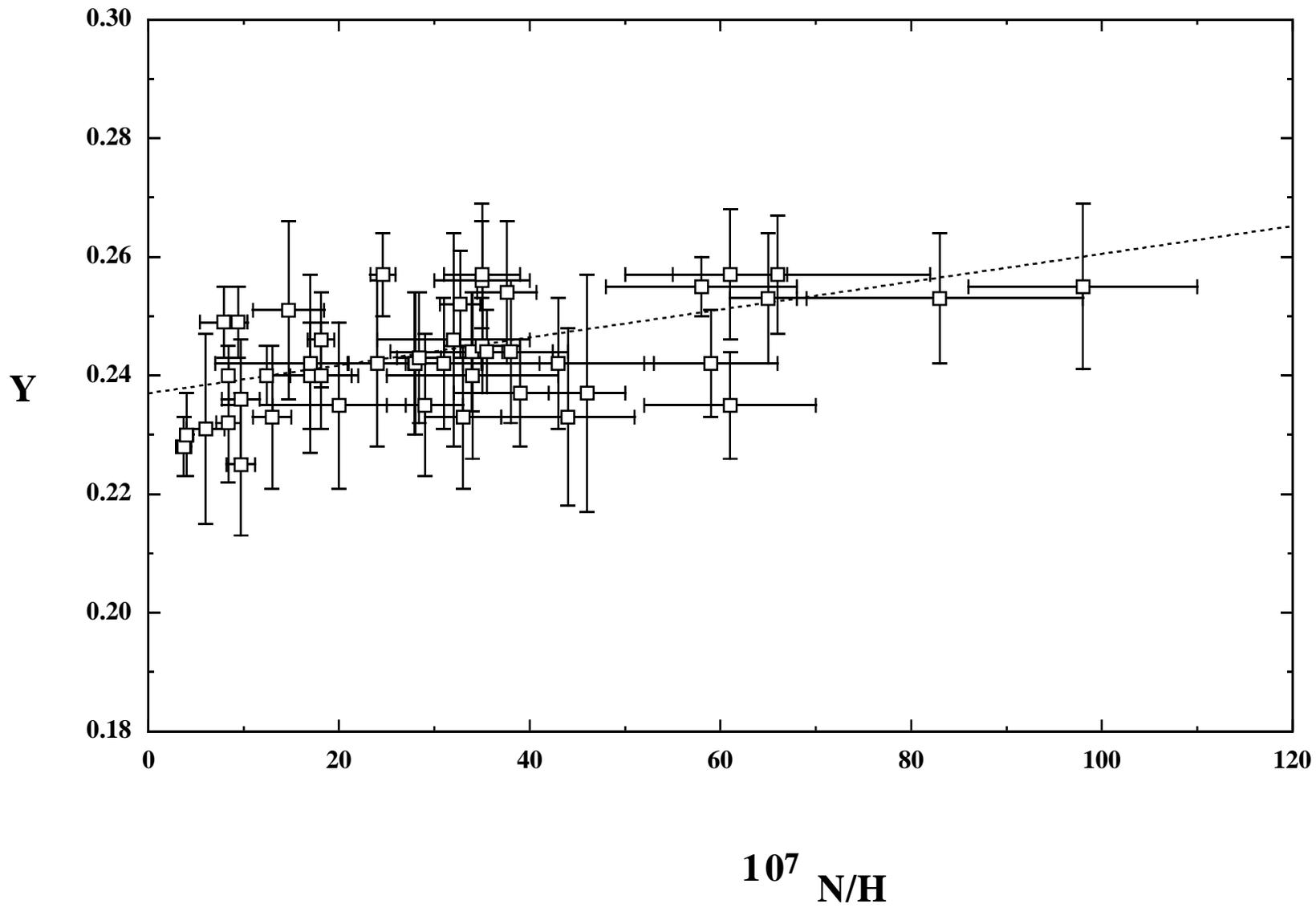
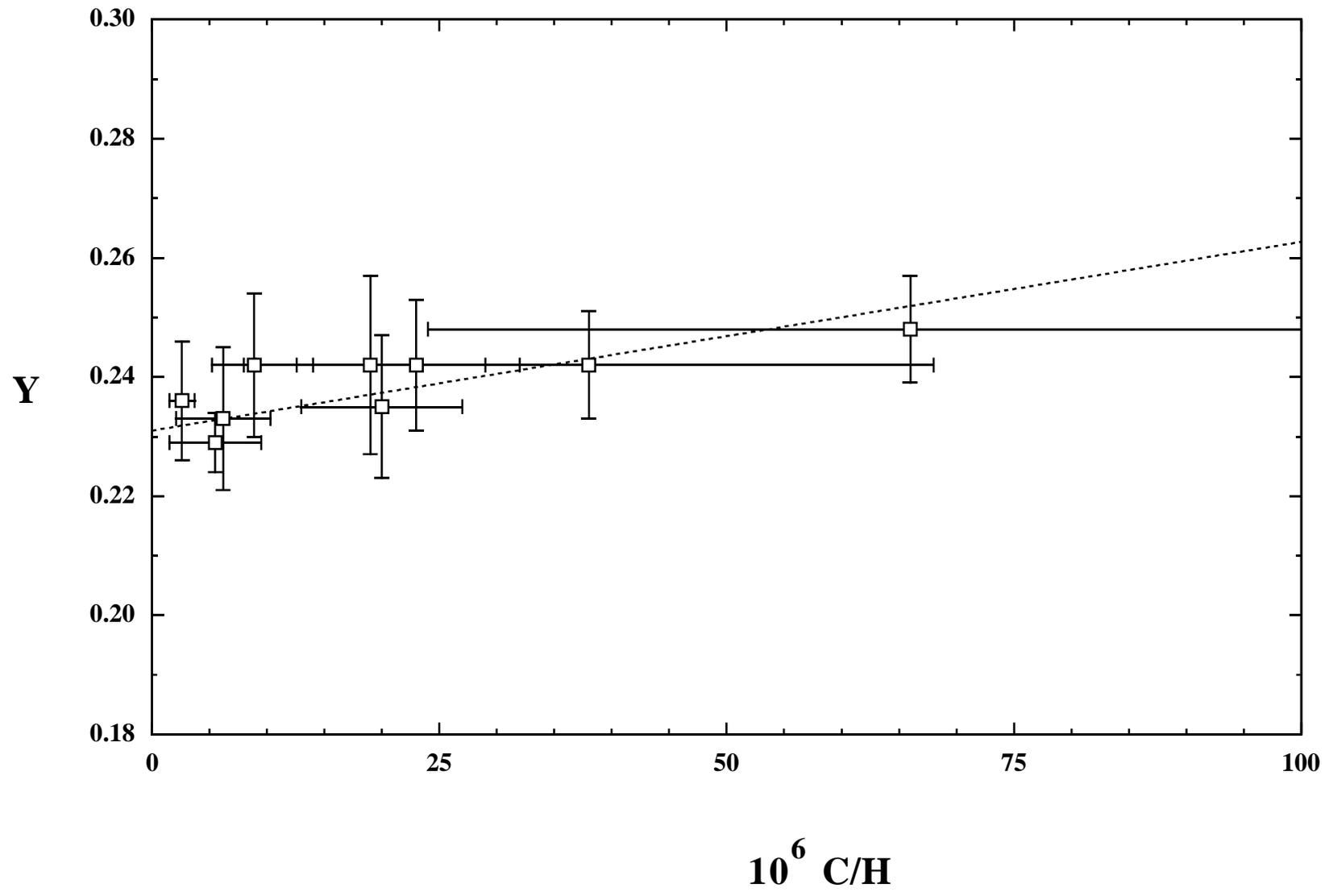
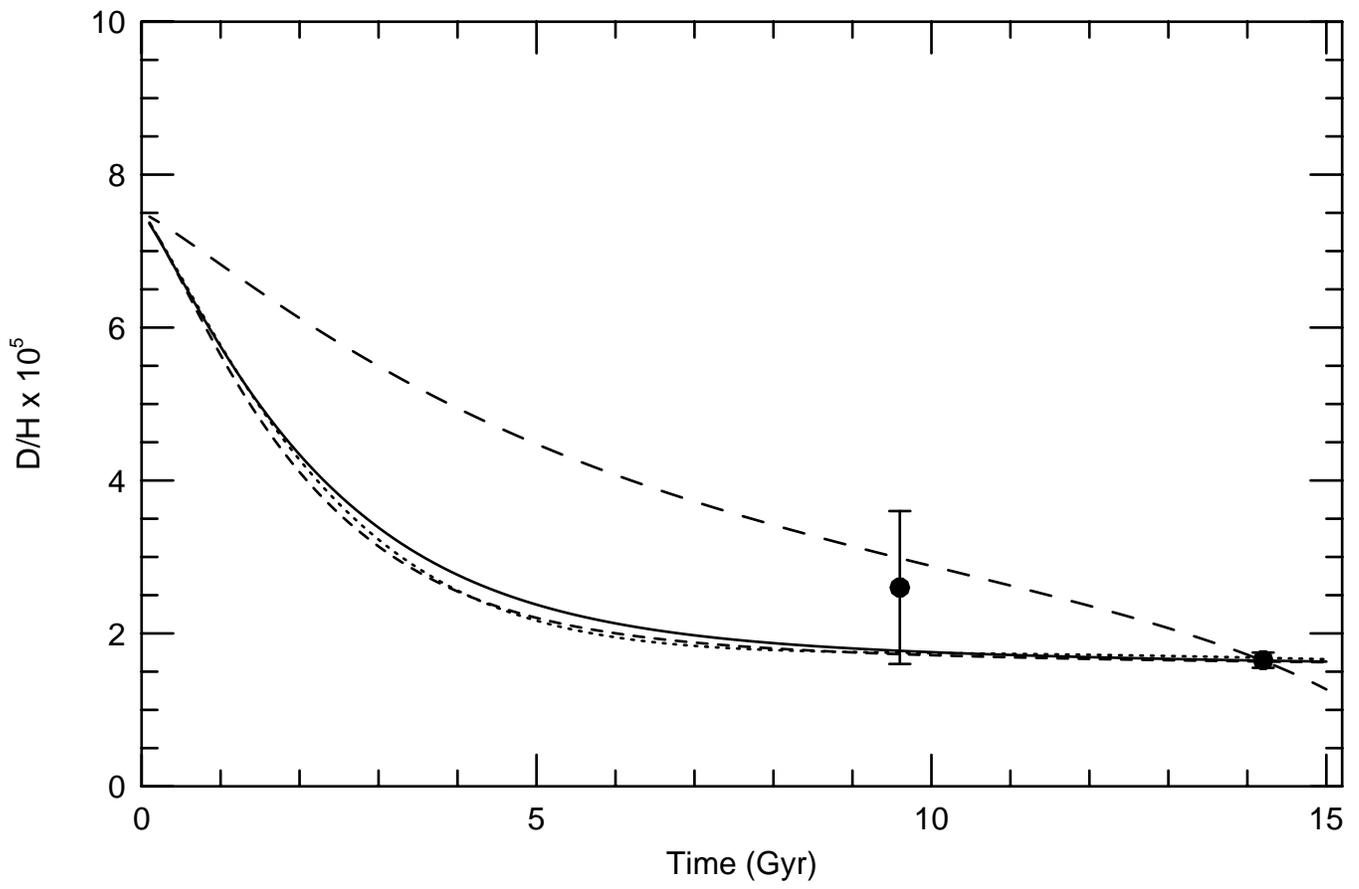
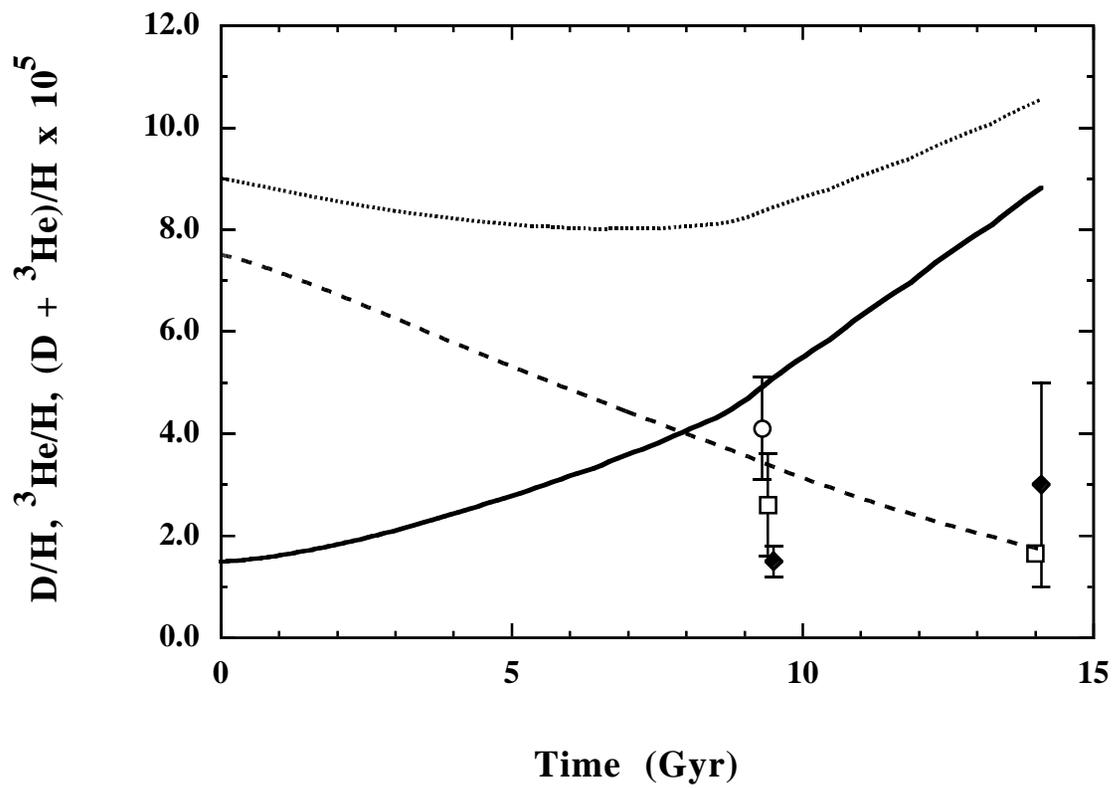
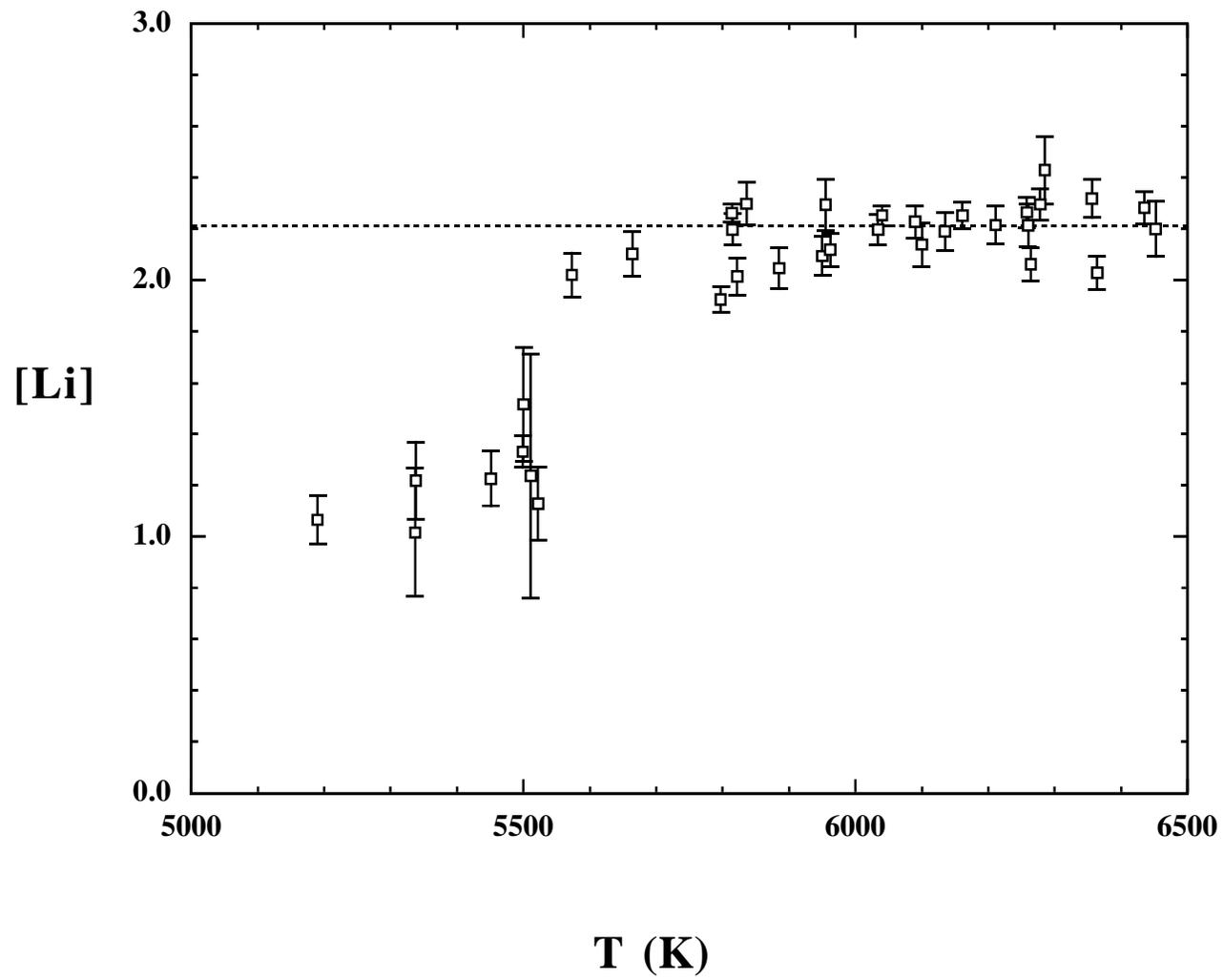
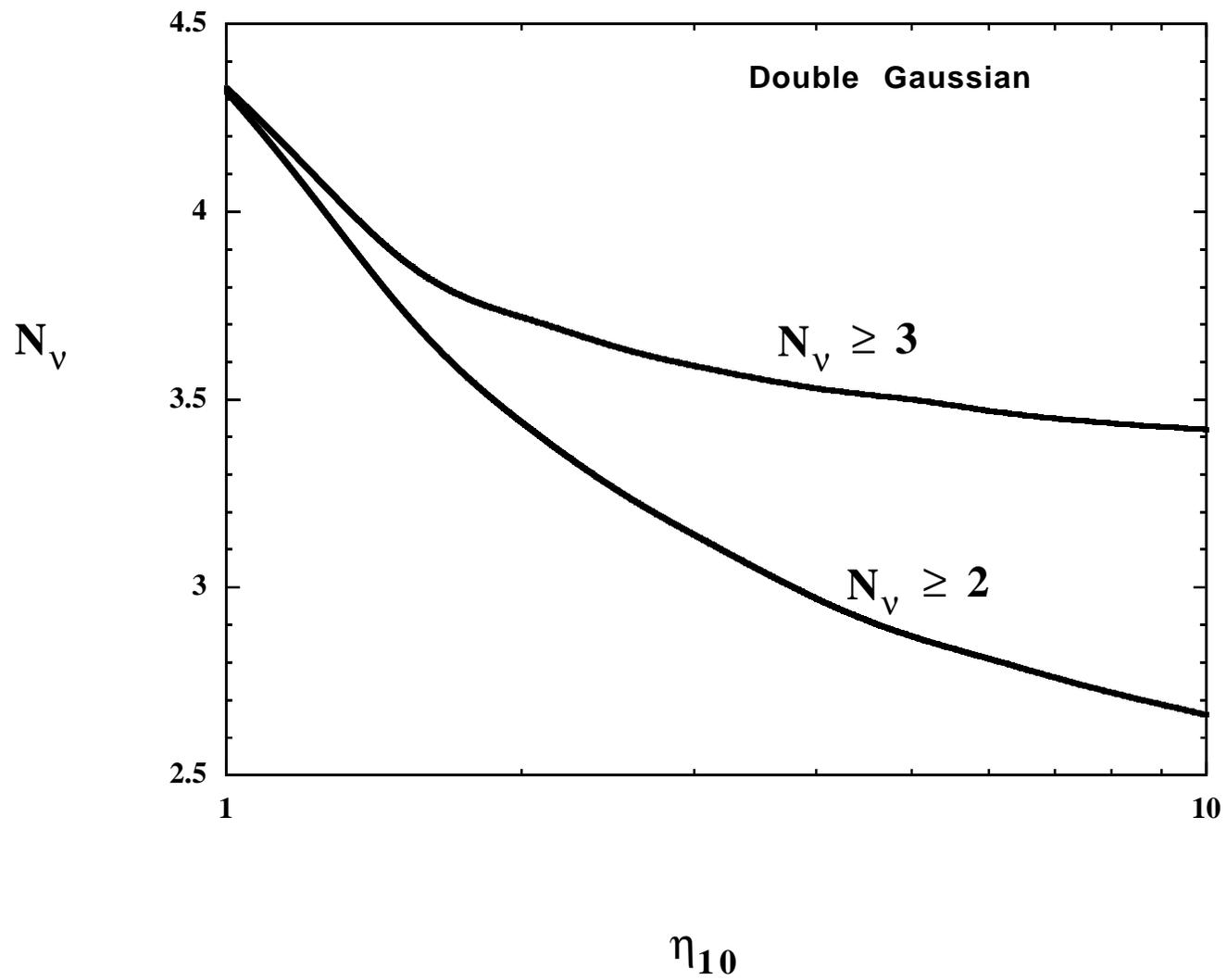
\end{document}